\documentclass[useAMS,usenatbib,usegraphicx]{mn2e}
\newcommand{\uv}{\mbox{$u$-$v$}}
\newcommand{\ex}[1]{\mbox{$\times 10^{#1}$}}

\newcommand{\kms}{\mbox{km s$^{-1}$}}
\newcommand{\Jb}{\mbox{Jy bm$^{-1}$}}
\newcommand{\alp}{\mbox{$\alpha^{\rm 1.4\, GHz}_{\rm 0.3 \, GHz}$}}
\newcommand{\acl}{\mbox{$\alpha^{\rm 4.9\, GHz}_{\rm 1.4 \, GHz}$}}
\newcommand{\Ra}[4]{\mbox{${#1}^{\rm h} \; {#2}^{\rm m} \; {#3}\fs{#4} $}}
\newcommand{\dec}[4]{\mbox{${#1}\degr \; {#2}\arcmin \; {#3}\farcs{#4} $}}
\newcommand{\GX}{\mbox{G21.5$-$0.9}}
\newcommand{\HI}{\mbox{H{\small\rmfamily{I}}}}
\newcommand{\HII}{\mbox{H{\small\rmfamily{II}}}}
\newcommand{\GY}{\mbox{G21.45$-$0.59}}
\newcommand{\GZ}{\mbox{G21.64$-$0.84}}
\newcommand{\PSRJ}{\mbox{PSR~J1833$-$1034}}
\newcommand{\Wmsr}[1]{\mbox{$\, \times ~10^{#1}$} {\mbox W~m$^{-2}$ Hz$^{-1}$ sr$^{-1}$}}
\newcommand{\lesssim}{\mbox{\raisebox{-0.3em}{$\stackrel{\textstyle <}{\sim}$}}}

\title[Radio Study of G21.5--0.9]{The Deepest Radio Study of the Pulsar Wind Nebula G21.5--0.9: Still No Evidence
for the Supernova Shell}
\author[Bietenholz et al]{M. F. Bietenholz$^{1,2}$, 
H. Matheson$^3$, S. Safi-Harb$^{3,4}$, C. Brogan$^4$, and N. Bartel$^2$ \\
$^1$Hartebeesthoek Radio Observatory, PO Box 443, Krugersdorp,
1740, South Africa \\
$^2$Department of Physics and Astronomy, York University, Toronto,
M3J~1P3, Ontario, Canada \\
$^3$Department of Physics and Astronomy, University of Manitoba, Winnipeg,
Manitoba, Canada \\
$^4$Canada Research Chair \\
$^5$NRAO, 520 Edgemont Rd, Charlottesville, VA, 22903, USA }

\begin{document}
 
\maketitle
\label{firstpage}

\begin{abstract}
We report on sensitive new 1.4-GHz VLA radio observations of the
pulsar wind nebula \GX, powered by \PSRJ, and its environs.  Our
observations were targeted at searching for the radio counterpart of
the shell-like structure seen surrounding the pulsar wind nebula in
X-rays.
Some such radio emission might be expected as the ejecta from the
$\lesssim$1000~yr old supernova expand and interact with the
surrounding medium.  We find, however, no radio emission from the
shell, and can place a conservative $3\sigma$ upper limit on its 1-GHz
surface brightness of 7~\Wmsr{-22}, comparable to the lowest limits
obtained for radio emission from shells around other pulsar-wind
nebulae.  Our widefield radio image also shows the presence of two
extended objects of low-surface brightness.  We re-examine previous
327-MHz images, on which both the new objects are visible.  We
identify the first, \GZ, as a new shell-type supernova
remnant, with a diameter of $\sim$13\arcmin\ and an unusual
double-shell structure.  The second, \GY, $\sim$1\arcmin\ in diameter,
is likely an \HII\ region.
\end{abstract}


\section{Introduction}
The crab-like supernova remnant \GX\ (SNR 021.5$-$00.9) harbours a
pulsar wind nebula (PWN), powered by the 61.86~ms pulsar, \PSRJ.
Despite having a high spin-down luminosity of 3.3\ex{37}~erg~s$^{-1}$
\PSRJ\ produces only faint pulsed emission, and was only recently
discovered \citep{Gupta+2005, Camilo+2006}.  Radio imaging of the PWN
spanning a $\sim$15-year period has shown that it is expanding quite
rapidly, having an expansion age of $870^{+200}_{-150}$~yr
\citep{G21.5expand}, making \PSRJ\ one of the youngest pulsars known.
Using a determination of the hydrogen column density, CO and \HI\
measurements, \citet{Camilo+2006} determined that the distance to \GX\
was $4.7 \pm 0.4$~kpc, which is consistent with earlier determinations
\citep[e.g.][]{Safi-Harb+2001, DavelaarSB1986}. We will adopt the value
of 4.7~kpc.

The PWN is bright in both the radio and X-ray, having luminosities of
$\sim$10\% and $\sim$1\% respectively of those of the Crab Nebula.

Recent radio images of \GX\ show a flat-spectrum PWN, with an angular
diameter of $\sim$1\arcmin\ \citep{G21.5expand}, which has filamentary
structure similar to that seen in the Crab Nebula.  The pulsar is
expected to have been born in a supernova explosion, and since it is
quite young, one might expect to see some emission associated with the
expanding shell of supernova ejecta.  However, despite a number of
studies in the radio \citep{BeckerK1976, WilsonW1976, BeckerS1981,
MorsiR1987, Furst+1988, Kassim1992, BockWD2001, G21.5expand}, no radio
emission from the supernova shock front has been seen.  In particular,
the 1-GHz surface brightness, $\Sigma_{\rm 1\, GHz}$, of any such
emission has been limited ($1\sigma$) to being $< 4$~\Wmsr{-21}
\citep{Slane+2000}.  Although this limit represents much brighter
emission than the faintest known supernova shell in our Galaxy,
G156.2+5.7, which has $\Sigma_{\rm 1 \, GHz} = 5.8$~\Wmsr{-23}
\citep{ReichFA1992}, \GX\ is much younger\footnote{For G156.2+5.7,
\citet{Yamauchi+1999} give an age of 15,000~yr and a distance of
1.3~kpc, but we note that both these values are uncertain by a factor
of $\sim$2.} and therefore might be expected to have brighter radio
emission assuming a similar ISM density and supernova explosion
energy.

\GX's centrally-condensed PWN is also seen clearly in X-rays
\citep{Slane+2000, Safi-Harb+2001, Warwick+2001, MathesonS2005,
Bocchino+2005, MathesonS2010}.  In contrast to the radio, however, a
fainter ``halo'' of X-ray emission, with a radius of $\sim$150\arcsec,
is indeed seen surrounding the PWN\@.  Both \citet{MathesonS2005} and
\citet{Bocchino+2005} argue that most of this halo X-ray emission is
not from the outer shock, but is rather due to dust scattering.
However, a relatively weak, limb-brightened X-ray component is also
seen on the eastern side, which has been interpreted as non-thermal
emission associated with the supernova shock \citep{Bocchino+2005,
MathesonS2010}.

With the goal of identifying any radio emission associated with this
non-thermal X-ray emission from the forward shock, we obtained
sensitive new observations of \GX\ and its surroundings.

\section{Observations and Data Reduction}
\label{sobs}

We observed \GX\ in the 1.4~GHz band on 2008 March 17, using the C
array configuration of the National Radio Astronomy
Observatory\footnote{The National Radio Astronomy Observatory, NRAO,
is a facility of the National Science Foundation operated under
cooperative agreement by Associated Universities, Inc.} (NRAO) VLA,
with a total time of 6~hours.  In this array configuration, the VLA is
sensitive to structures up to 15\arcmin\ in size, so the structure of
the putative radio-shell around \GX\ should be well-sampled.  In order
to maximise the field-of-view, \uv~coverage and dynamic range, we
observed in spectral line mode using spaced centre frequencies of
1.4649 and 1.3851~GHz in the two intermediate-frequency (IF) channels.
We phase-referenced our observations to the compact source PMN
J1832$-$1035,
whose position is accurate to better than 1\arcsec,
therefore our astrometric uncertainty is dominated by contributions
from noise and errors in phase-referencing.  We estimate our
astrometric uncertainty at $<4$\arcsec.
Our flux density scale was set from observations of 3C~48 and 3C~286.
The data reduction was carried out using standard procedures from
NRAO's AIPS software package.

Our final images were made from self-calibrated visibility data,
using CLEAN deconvolution with multiple non-coplanar facets.  Since
our two IF frequencies differ by $\sim$6\%, sources with different
spectral indices could show noticeable differences in relative
brightness between the two IFs.  In particular, in \citet{G21.5expand}
we found that the spectral index, $\alpha$ (where the flux density $S$
at frequency $\nu$ is $\propto \nu^\alpha$), of the \GX\ PWN is quite
uniform over the nebula, with a value of $+0.08^{+0.06}_{-0.09}$.  In
contrast to a PWN, a supernova shell would be expected to have a
notably steeper spectrum, with a typical value of $-0.8 \sim -0.4$
\citep{Green2009}.

In order that such spectral-index differences not limit the dynamic
range in the deconvolved image, we chose to image and deconvolve our
two IFs separately.  We then averaged the resulting two images to
obtain our final, combined image.  The final image should correctly
represent the brightness at the mean frequency of 1.43~GHz regardless
of the spectral index.  Separate imaging of the two IFs did in fact
produce a slightly higher dynamic range than did combining the two IFs
before the Fourier transform stage.  

We also discuss 327~MHz images of this region resulting from earlier
VLA observations of a large segment of the Galactic plane.  These
327-MHz observations were taken as part of an effort to survey the
inner Galactic plane at low radio frequencies, and are described in
\citet{Brogan+2006, Brogan+2005}.  The FWHM resolution is 85\arcsec. A
lower-resolution image made from these data, which covers the part of
the plane relevant to this paper, is reproduced in
\citet{Brogan+2005}.

Finally we also discuss a \emph{Chandra} X-ray image of \GX\ and its
environs, produced by combining a total of $\sim$520~ksec of
\emph{Chandra} observations in the energy range 0.2 to 10 keV\@.  The
X-ray data are described in \citet{MathesonS2005, MathesonS2010}.

\section{Results: 1.4-GHz Wide-Field Image}
\label{simg}

\begin{figure*}
\centering
\includegraphics[width=0.9\textwidth]{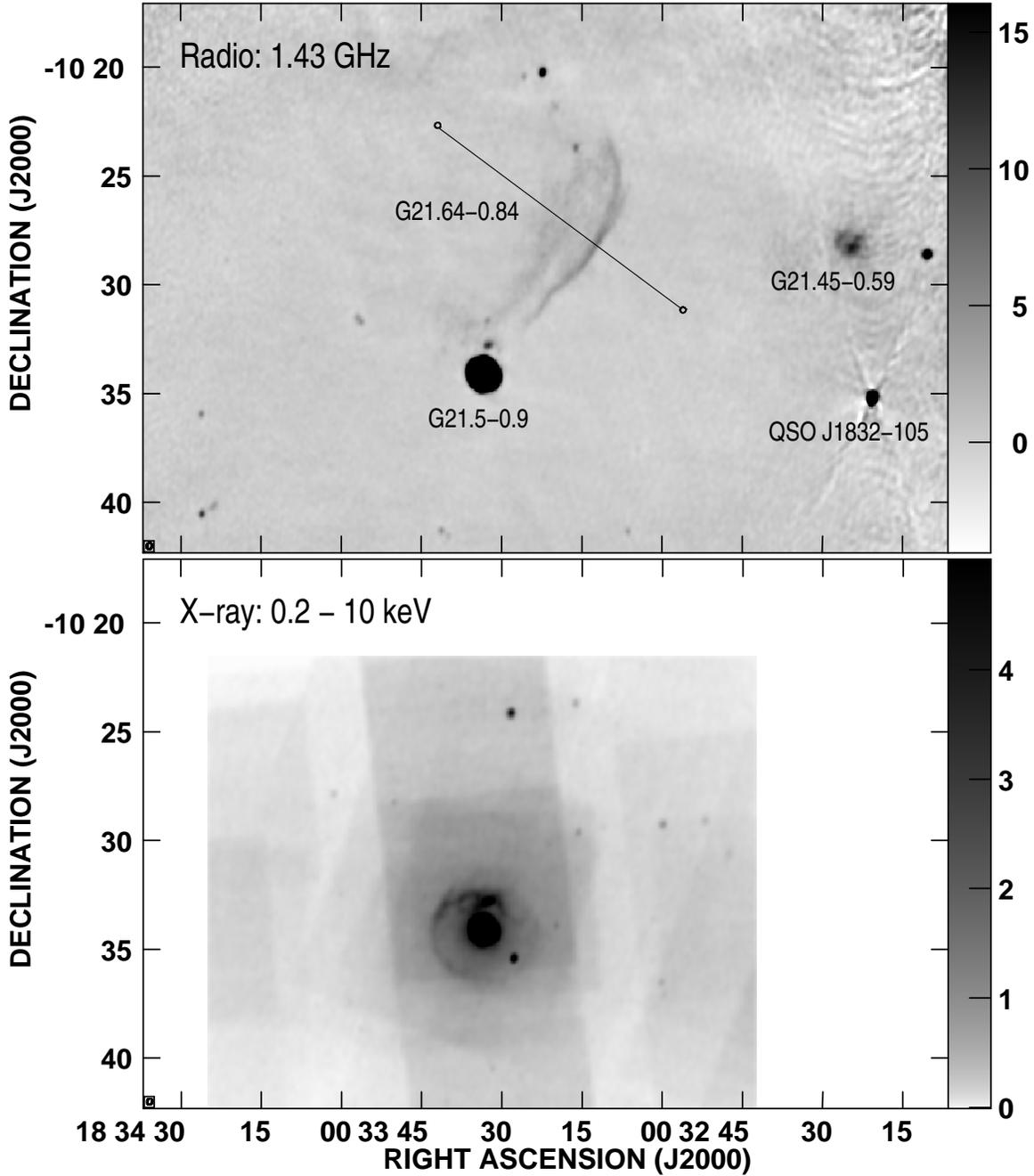}
\caption{{\bf Top:} a wide-field 1.4-GHz radio image of the region
around \GX\@.  The greyscale range of $-2$ to 10~m\Jb\ was chosen to
emphasise any weaker background emission, therefore saturating the
\GX\ PWN itself, which had a peak brightness of 621~m\Jb.  The sources
we discuss in the text are labelled.  The image has been corrected for
the VLA primary beam response, with the result that the rms is $\sim
260 \; \mu$\Jb\ near \GX, which was the pointing centre, and increases
with increasing distance away from it.  The FWHM of the convolving
beam was $18.1\arcsec \times 13.8\arcsec$ at p.a.\ $-12$\degr.
Low-level artifacts are visible in the vicinity of the bright source,
QSO~J1832-105.  The diagonal line segment shows the location and
orientation of the slice through \GZ\ shown in Fig.~\ref{fslice}.
{\bf Bottom:} An X-ray image of \GX, showing the same field of view,
and convolved to the same resolution, as the radio image above, made
by combining a total of $\sim$520~ksec of \emph{Chandra} observations
in the energy range 0.2 to 10 keV \citep[for details,
see][]{MathesonS2010}.  The greyscale is chosen so that the \GX\ PWN
in the centre is saturated in order to show the fainter X-ray emission
surrounding it. Different regions of the image have been observed with
different effective exposure times, resulting in a varying background
level.  The bright point source in the south-western part of the X-ray
halo is an unrelated emission-line star, SS~397.}
\label{fwide}
\end{figure*}
We show the wide-field radio-image of the \GX\ region in the top part
of Fig.~\ref{fwide}.  This image represents only part of the imaged
area, chosen to show the sources of interest.  We further chose the
range of greyscale in order to show the weaker sources, with the
result that the stronger ones such as \GX\ itself are saturated.
Clearly visible in this image is the feature called the ``northern
knot'', $\sim$2\arcmin\ to the north of the centre of \GX\@.  This
image is corrected for the response of the primary beam (which has a
FWHM of 31\arcmin\ at 1.43~GHz).  As a consequence the rms background
level increases with distance from the pointing centre.  Near \GX, the
rms background brightness was $\sim260\,\mu$\Jb.  This is the deepest
radio image
so far obtained of \GX\ and its surroundings.  The bottom part of
Fig.~\ref{fwide} shows the \emph{Chandra} X-ray image for the same
field.

In addition to \GX\ there are a number of other sources visible.  The
brightest is QSO~J1832-105, with a 1.43-GHz flux density of 1.07~Jy.
Two other resolved sources are visible at approximately RA = $18^{\rm
h}$~33\farcm 2, decl.\ = $-10$\degr\ 27\farcm 0 and RA = $18^{\rm
h}$~32\farcm 4, decl.\ = $-10$\degr\ 28\farcm 3.
We call these sources \GZ\ and \GY, as they are likely both Galactic,
and we will discuss them below.  There are also a number of weaker
unresolved sources visible, which are likely extragalactic and which
we do not discuss further.  

Also present in the full image and included in the deconvolution and
self-calibration was the supernova remnant Kes~69, which is to the
northwest of \GX.  We choose to exclude Kes~69 from the portion of the
image displayed in Fig.~\ref{fwide}, as it is beyond the 25\% point of
the primary beam, and
the image details are not reliable \citep[see][for a radio image of
Kes~69]{Kassim1992}. Some artifacts are visible in the western and
northwester parts of the image, due to J1832-105 and Kes~69.

\subsection{\GX}
\label{sg21obs}

\begin{figure}
\centering
\includegraphics[width=0.49\textwidth]{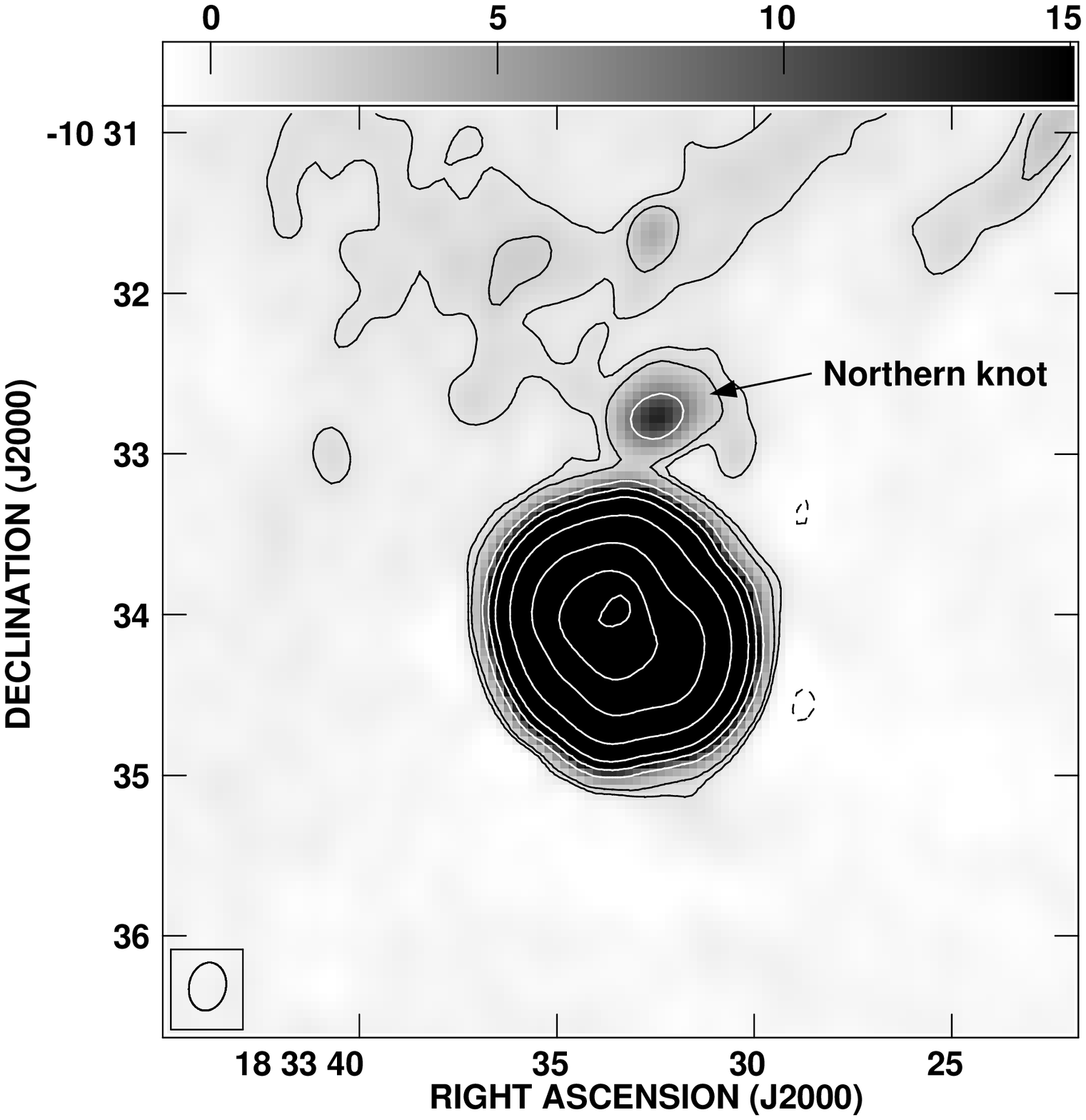}
\caption{A 1.4-GHz image of \GX.  The figure represents a detail of
the top panel of Fig.~\ref{fwide}. The contours are drawn at $-1$,
1, 2, 4, 8, 16, 32, 100,300, 500 and 600 m\Jb.  The greyscale is
chosen to emphasise the fainter emission, and is labelled in m\Jb. The
convolving beam FWHM, indicated at lower left, was $18.8\arcsec \times
13.8\arcsec$ at p.a.\ $-12$\degr.  The peak brightness in this sub-image
was 621~m\Jb, and the rms background noise was 260~$\mu$\Jb.  The
``northern knot'' is labelled.  Also visible at the top of the image
is the lower portion of \GZ.}
\label{fg21only}
\end{figure}

In Fig.~\ref{fg21only} we show a detail of the radio image showing the
\GX\ PWN\@. The image corresponds well, albeit at lower resolution, to
the one of \citet{G21.5expand}.  \GX\ had a total 1.43-GHz flux
density of $7.0 \pm 0.4$~Jy,
with a peak surface brightness at our resolution of $0.62 \pm
0.03$~\Jb, corresponding to a brightness temperature of $1500\pm75$~K
(where the uncertainties are dominated by the assumed 5\% uncertainty
in the flux-density calibration). 

On our 327~MHz image \citep[image not reproduced here;
see][]{Brogan+2005}, \GX\ is also clearly visible, although only
marginally resolved.  Comparison of the flux density of $7.3 \pm
0.7$~Jy determined from the 327-MHz image with that at 1.4~GHz yields
an integrated spectral index between these frequencies, $\alp$, of
$0.0 \pm 0.1$, consistent with other determinations of \GX's radio
spectral index \citep[e.g.,][]{G21.5expand}.

There is a relatively compact radio source, approximately 1\arcmin\ to
the north of the main body of \GX, which corresponds to the X-ray
feature called the ``northern knot''\footnote{This feature was also
called the ``northern spur'' by \citet{Bocchino+2005}}
\citep[e.g.,][]{MathesonS2010}. B. Gaensler, in a private
communication, reported seeing the northern knot in earlier 1.4-GHz
radio observations.  The northern knot has a total 1.43-GHz flux
density of $20.2 \pm 1.8$~mJy, and is marginally resolved in our
image, with an intrinsic FWHM size, as estimated by an elliptical
Gaussian fit, of $18\arcsec \times 8\arcsec$ at p.a.\ 105\degr.  The
peak brightness position in the radio is consistent with that seen in
the X-ray when the latter is convolved to the same resolution,
as can be seen in the top and bottom panels of Fig.~\ref{fwide}.

The northern knot is not discernible on our lower-resolution 327~MHz
image, being blended with \GX\ itself.  The knot is marginally visible
in an image made from the 5-GHz data of
\citet{G21.5expand}\footnote{Note that the northern knot is not within
the portion of the image reproduced in \citet{G21.5expand}. We have
re-imaged the visibility data to include the knot region.}.  We
estimate a total 5-GHz flux density for the northern knot of $19 \pm
7$~mJy, resulting a nominal value for the knots spectral index, \acl,
of $-0.04$, similar to that body of \GX\@.  However, the uncertainties
on the spectral index are large, and the $p = 2.3\% \; (2\sigma)$
limits on \acl\ are $-1.3$ and $+0.5$.

What is notably absent from our deep 1.4-GHz image is radio emission
corresponding to the limb-brightened X-ray emission visible in the
bottom panel of Fig.~\ref{fwide}.  Although some diffuse radio
emission appears north of the northern knot, we believe that it is not
the counterpart of the X-ray shell of G21.5$-$0.9, but more likely
associated with G21.64$-$0.84, which we discuss below. We derive a new
upper limit to the radio emission from the SNR shell in \S~4.1.

\subsection{\GZ}
\label{sstreamerobs}

A so far uncatalogued source, \GZ, is distinctly visible in the
1.4-GHz image in Fig.~\ref{fwide}.  It is a large elongated
structure, which extends from near \GX\ to the north-northwest.  It is
brightest in the middle, reaching a peak 1.43-GHz brightness of $6.8
\pm 0.5$~m\Jb\ (corresponding to a brightness temperature of $17 \pm
1$~K)\@. It has a total length of $\sim$10\arcmin.  It seems to
consist of two relatively distinct and roughly parallel curved ridges
of emission, with the western or outer one being better defined,
and the eastern or inner one being more diffuse and $\sim$1.5\arcmin\
distant.  The total 1.43-GHz flux density in the \GZ\ is $660 \pm
50$~mJy, with each of the two ridges having approximately half the
total.

\begin{figure}
\centering
\includegraphics[width=0.49\textwidth]{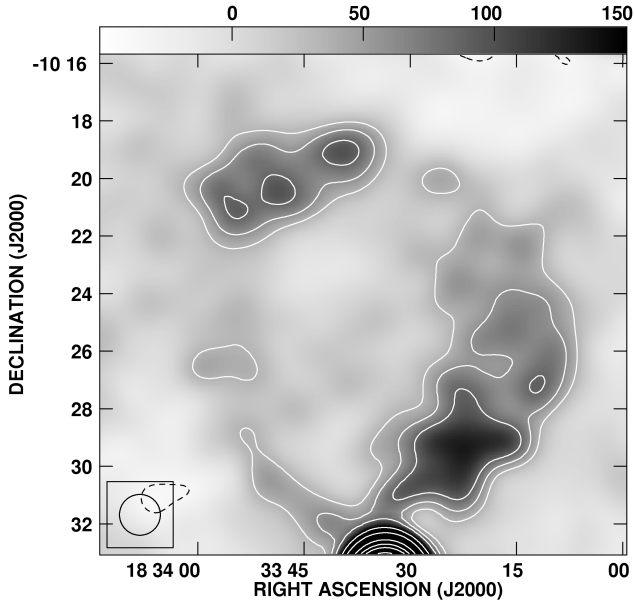}
\caption{A 327~MHz image of \GZ.  The FWHM resolution was 85\arcsec,
and is indicated at lower left.  The greyscale is labelled in m\Jb,
and the rms background was $\sim$14~m\Jb. The contours are drawn at
$-0.6$, 0.6, 1, 1.5, 3.0, 5, 10, 15, 20, and 30\% of the peak
brightness which was 5.73~\Jb. \GX\ is partly visible at the bottom of
the image.}
\label{fstreamerp}
\end{figure}

\GZ\ is also visible in the 327~MHz image, and we show a detail in
Fig.~\ref{fstreamerp}.  At this frequency, due to the larger primary
beam, much more of the source is visible than at 1.4~GHz, and it can
bee seen to be a relatively circular shell-structure..  The centre of
the shell is at RA = 18$^{\rm h}$ 33\farcm6 and decl.\ $-10$\degr\
25\farcm4, or at $l = 21.64, b = -0.84$.  Our name for this source,
\GZ, is based on the centre position of the shell as seen at 327~MHz.
The outer angular diameter of the shell is $\sim$13\arcmin.
The total flux density at 327~MHz is 2.8~Jy, with an estimated
uncertainty of $\sim$0.5~Jy due to the somewhat uncertain zero-level
in the images.  The 327-MHz flux density of the part visible in the
1.4~GHz image is $1.4 \pm 0.2$~mJy, and over this region the
integrated value of \alp\ is $-0.5 \pm 0.1$.  A somewhat steeper
spectrum is suggested for the northern and eastern sides of the shell,
which are not visible at 1.4~GHz, however due to the uncertain
zero-levels and low signal-to-noise ratios, a constant value of $\alp
\sim -0.6$ for the whole shell is probably not excluded.
The averaged over the whole of the 327-MHz surface brightness is
2.5~\Wmsr{-21}, and if we take $\alpha = -0.5$, then $\Sigma_{\rm
1\,GHz} = 1.4$~\Wmsr{-21}.

We include the \GZ\ region in the wide-field X-ray image shown in the
lower panel of Fig.~\ref{fwide} above.  Although this region was not
the primary target, and the instrumental background differs in different
parts of \GZ, no sign of X-ray emission which might be associated with
\GZ\ is seen down to limits of $10^{-3}$ times the peak brightness of
\GX.  At the radio peak-brightness position of \GZ, we estimate that the
X-ray exposure was $\sim$130~ks.

\begin{figure*}
\centering
\includegraphics[width=0.8\textwidth]{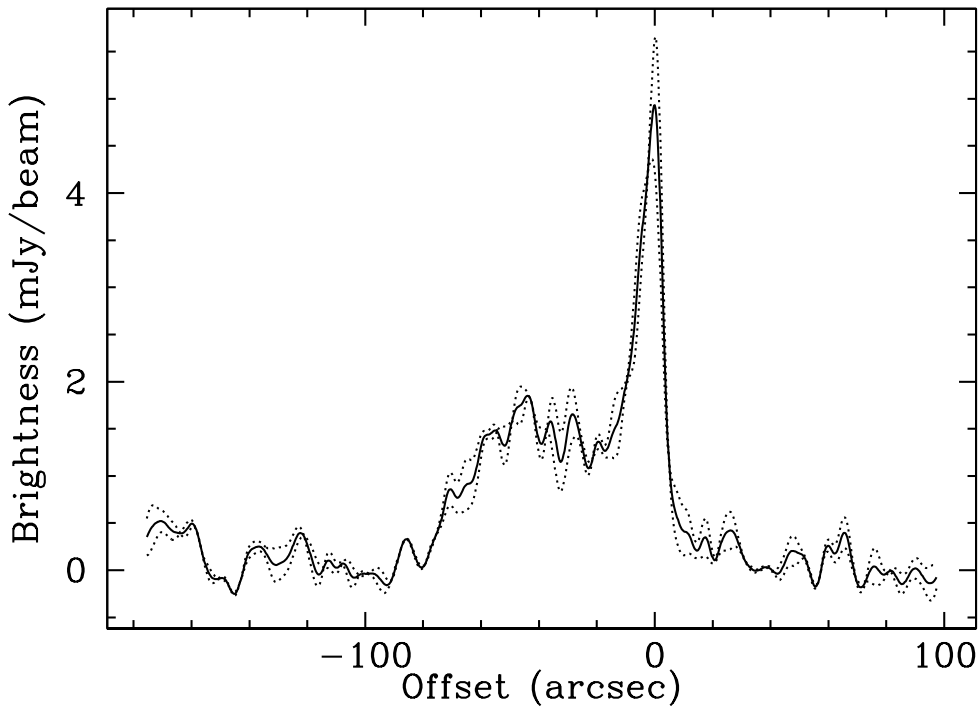}
\caption{A profile drawn through \GZ, at p.a.\ = $-127$\degr\ taken
from the image shown in Fig.~\ref{fwide}, where the location of the
slice is also shown.  The offsets are in arc-seconds, measured from
the slice peak, with a positive offset being to the southwest.  To
increase the signal-to-noise, we averaged three parallel slices, each
displaced by $\pm 7\arcsec$ transverse to the slice direction.  The
solid line shows this average value, while the dotted lines indicate
the rms over the three slices.  The slice peak is 5.03~m\Jb, and it is
located at \Ra{18}{33}{12}{}, \dec{-10}{28}{09}{3}.  The half-power
points occur at displacements of $-20$\arcsec\ and +11\arcsec\ from
this peak.}
\label{fslice}
\end{figure*}

In Fig.~\ref{fslice}, we show a profile through \GZ\ centred at
\Ra{18}{33}{12}{}, \dec{-10}{28}{09}{3}, which is near its peak
brightness position (the location of the profile is indicated in Fig.
\ref{fwide}).  The profile through the brighter ridge is asymmetric,
with the north-east side being approximately half as steep as the
southwest one.  In particular, along the slice direction, the
half-maximum point in the northeast direction is reached at a distance
of 20\arcsec\ from the peak, whereas in the southwest direction it is
reached at a distance of 11\arcsec.  The FWHM on the slice is
therefore 31\arcsec, which is notably larger than our resolution of
$\sim$13\arcsec\ along the slice direction.  The intrinsic FWHM is
therefore close to the observed one of 31\arcsec. The more diffuse,
fainter ridge has a FWHM of $\sim$3\arcmin.
We discuss the nature of \GZ\ below in \S~\ref{sstreamerconc}.

\subsection{\GY}  
\label{sgyobs}

A further weak resolved source is visible in both our 1.4-GHz
(Fig.~\ref{fwide}) and 327-MHz images near RA 18$^{\rm h}\; 32^{\rm
m}$, decl.\ = $-10\degr$ 28\arcmin\ ($l = 21.459, b = -0.59$), which
we call \GY\@.  It is approximately 1\farcm5 in diameter. Its
327-MHz peak brightness was $209\pm24$~m\Jb\ (at 85\arcsec\
resolution), corresponding to a brightness temperature of $330 \pm
38$~K\@.

We estimate total flux densities for \GY\ of $\sim$0.3~Jy at 1.4~GHz
and $\sim$0.5~Jy at 327~MHz, corresponding to \alp\ $\sim -0.3$,
although this spectral index estimate should be treated with caution
as the total flux densities are rather uncertain because of the
uncertain extent of the source and zero-levels, as well as the large
primary-beam correction factor of $\sim$2.6 at 1.4-GHz.

\GY\ has likely been identified already in the 2.7-GHz continuum
survey of the Galactic plane by \citet{Reich+1984}, who list the
source RFS~323 at a position about 90\arcsec\ away from our peak
brightness position for \GY, and having a total 2.7-GHz flux density
of $0.50\pm0.15$~Jy.
The position offset between \GY\ and RFS~232 is larger than
either our own astrometric uncertainty of $<4$\arcsec\ or that of
20\arcsec\ listed by \citet{Reich+1984}.  However, the FWHM resolution
of the Reich et.\ al survey was 4.3\arcmin, so the positional offset
is only approximately 1/3 the Reich et al.\ beamwidth, and thus
probably not significant.  We consider that RFS~232 can thus be
identified with \GY, with Reich et al.'s 2.7-GHz flux density implying
$\alpha^{\rm 2.7\, GHz}_{\rm 0.3 \, GHz} \sim 0$.

We note that \GY\ is also visible in the 1.4-GHz VLA Galactic Plane
Survey \citep[VGPS][]{Stil+2006}, and has a peak brightness position
consistent to within 2\arcsec\ with the one we measured at that
frequency. We consider our peak brightness position for \GY\ more
reliable than that listed for RFS~232 by \citet{Reich+1984}.

Our X-ray observations do not extend to \GY, so its X-ray brightness
is unknown.  We will discuss the nature of \GY\ in \S~\ref{sgyconc}
below.

\section{Discussion}
\label{sconclude}

\subsection{\GX\ and Limits on Radio Emission from its Forward Shock}
\label{sshock}

Our 1.4-GHz radio image of the \GX\ region (Fig.~\ref{fwide}, top) is
dominated by the pulsar-wind nebula \GX.  Our observations are
sensitive to structures much larger than the \GX\ PWN, so there is no
reason to suspect any significant deficit in the recovered flux
density due to missing short interferometer spacings. Indeed, our
total recovered flux density of $7.0\pm0.4$~Jy is consistent within
the uncertainties with the value measured by \citet{Altenhoff+1970}
and the radio spectrum compiled by \citet{Salter+1989}.

Except for the northern knot, no radio emission is visible which can
be associated with the X-ray halo or the limb-brightened feature seen
in X-ray.  Conservatively scaling the rms background brightness of our
image (260~$\mu$\Jb\ at 1.43~GHz) to the standard frequency of 1~GHz
using a steep assumed spectral index\footnote{To determine the limit,
we use $\alpha = -0.8$ as the steepest reasonable spectrum for a
shell. As the difference between the observed frequency of 1.43~GHz
and the nominal one is not large, our result is not sensitive to
reasonable changes in $\alpha$, for example, using a more usual value
of $\alpha = -0.5$ would lower our $3\sigma$ surface brightness limit
by $\sim$8\% to 1.4~\Wmsr{-21}.}
we obtain a $3\sigma$ upper limit on $\Sigma_{\rm \, 1 GHz}$ of
1.6~\Wmsr{-21}.

Since the X-ray limb detected at a radius of $\sim$150\arcsec\ likely
indicates the location of the forward shock
\citep[see][]{Bocchino+2005, MathesonS2010}, we can place more
stringent limits on the radio brightness of the shell by integrating
over the shell region.  Over an annular region with an outer angular
radius of 150\arcsec\ and a thickness of 20\% of the outer radius, the
observed total 1.43-GHz flux density is $12 \pm 6$~mJy,
with a $3\sigma$ upper limit of 30~mJy, corresponding to an average
1.43-GHz spectral luminosity of $< 8\ex{20}$~erg~s$^{-1}$~Hz$^{-1} (D
/ 4.7 {\rm kpc})$
or a surface brightness of $< 1.2 \; \mu$Jy per square arc-second.
Again conservatively scaling to 1~GHz with $\alpha = -0.8$ results in
a $3\sigma$ limit on the average surface of $\Sigma_{\rm 1 \, GHz} <
7$~\Wmsr{-22}.  Our assumption for the shell geometry is conservative:
for shells thicker than 20\%, as might be expected for an age of
870~yr \citep[see, e.g.,][]{JunN1996a, Chevalier1982c}, the
surface-brightness limit would be lower.  In addition, as noted above,
for values of $\alpha$ flatter than the assumed $-0.8$, the limit
would also be slightly lower, with $\alpha = -0.5$ resulting in a
value of 6~\Wmsr{-22} for the limit.  Finally, the annular region
above includes real emission due to \GZ, which is probably not
associated with \GX.  If we exclude this contribution,
the limit on the average surface brightness decreases by approximately
20\%.

The limit on the radio surface brightness of \GX's shell, which
corresponds to $<6\ex{-4}$ times the peak surface brightness of the
PWN, is $\sim$17 times lower than those previously obtained for \GX\
\citep{Slane+2000}.

It is expected that the shocks formed as the shell of supernova ejecta
plough through the interstellar medium would both accelerate particles
and amplify the ambient magnetic field by compression and through
instabilities.  Radio emission from this region might therefore be
expected, especially for a remnant as young as \GX.  Where then is
this radio emission?  

We note first that \GX\ is not unique in having no detectable radio
emission from the supernova shell.  In fact, most other young PWNe
seem to show little radio emission from the putative shell, with the
best studied example being the Crab Nebula, which is one of the
youngest known and also one of the very few identified with a
supernova (SN~1054), and which does not have any radio emission from
the ejecta shell down to a $3\sigma$ upper limit of 7.5~\Wmsr{-22}
\citep{Frail+1995}.
3C~58, also a young PWN, also has no detectable radio emission from
the shell to a $3\sigma$ surface brightness limit of $\lesssim \,
5$\Wmsr{-22} \citep{ReynoldsA1985,3C58-2001}.
In other words, the present limits on the radio surface brightness of
\GX' shell, although low, are not unusual when compared to those for
other bright PWNe.
What is unusual about \GX, however, is that unlike either the Crab or
3C~58\footnote{We note that a shell of thermal X-ray emission was 
seen in 3C~58 by \citet{GotthelfHN2007}.  The diameter of $\sim$5.6~pc
of this shell is smaller than the maximum extent of the PWN, which is
$\sim$8.5~pc.  The thermal X-ray emission is therefore likely
associated with supernova ejecta swept up by the expanding PWN rather
than the original forward shock from the supernova.},
X-ray emission from which can reasonably be attributed to the forward
shock \emph{has} been detected \citep{Bocchino+2005, MathesonS2010}.
\citet{MathesonS2010} fit the X-ray spectrum of the limb after
subtracting the halo emission, in other words the non-thermal part of
the X-ray emission thought to be due to the forward shock, with
broad-band models consisting of a power-law with an exponential
high-energy cutoff \citep[model \emph{srcut, }][]{Reynolds1998}.
Using an assumed radio spectral index of $\alpha = -0.5$, their best
fit model would imply $\Sigma_{\rm 1 \, GHz} = $2.3~\Wmsr{-22}, which
is well below our measured $3\sigma$ limit of 7~\Wmsr{-22}
The absence of radio emission from \GX's forward shock is therefore
consistent with the observed non-thermal X-ray emission.

Although we detect no radio emission from the shell, we do clearly
detect \GX's northern knot in our deep 1.4-GHz image.  A marginal
detection at 4.8~GHz results in a nominal spectral index, \acl, of
$\sim$0, albeit with large uncertainties.  This nominal value of \acl\
is similar to that for the body of \GX, suggesting that the knot, like
the body of \GX, consists of electrons energised by the pulsar.
However, the alternate hypothesis of the knot consisting of electrons
accelerated in the forward shock and having \acl\ $\sim -0.5$ is not
excluded by the data, so no firm conclusions as to its nature
can be drawn..

\subsection{\GZ: A New Shell Supernova Remnant}
\label{sstreamerconc}

We have identified a new shell-like radio source in our radio images,
which we have called \GZ.  In the high-resolution 1.4-GHz image
(Fig.~\ref{fwide} top), only the western part of the shell is
visible, but most of the shell can be seen in the 327~MHz image
(Fig.~\ref{fstreamerp}).  Its radio spectral index of $\sim -0.6$
(\S\ \ref{sstreamerobs}) is consistent with those seen for shell-type
SNRs \citep[see, e.g.,][]{Green2009}.

We examined the $8 \, \mu$m infra-red images from the Spitzer Glimpse
survey \citep{Benjamin+2003}, and find no infra-red emission for
\GZ, which implies that \GZ\ is unlikely to be either an \HII\ region
or a wind blown bubble, which latter often have a shell-like structure
but are almost inevitably accompanied by infra-red emission
\citep[e.g.,][]{Brogan+2006}.  It seems most probable that \GZ\ is an
as-yet uncatalogued supernova remnant.

We noted in \S~\ref{sstreamerobs} above that no X-ray emission is seen
from \GZ\ to levels of $\sim$0.1\% of the peak brightness of \GX.  The
lack of X-ray detection, however, does not argue against the supernova
remnant hypothesis: \citet{Green2009} has a comprehensive list of
Galactic SNRs and finds that only $\sim$40\% of them are detected in
X-ray, with high absorption in the X-ray being seen for many.

Could \GZ\ represent the radio emission associated with the forward
shock in \GX, the searched for which motivated the observations?  We
think this possibility very unlikely, as the centre off the new shell
is clearly displaced from the \GX\ PWN\@.  If \PSRJ\ had been at the
center of the newly-identified radio shell 870~yr ago, then its
average speed since then must have $\sim$15,000~\kms, which is far
higher than the speeds seen for pulsars. Moreover, X-ray emission from
\GX's forward shock has in fact been identified, and is not coincident
with \GZ\@.  We therefore identify \GZ\ as a previously unidentified
shell-type SNR, which is unrelated to \GX.

SNR shells typically have steeper brightness gradients at their
outside edges than the inside ones, thus naturally explaining the
asymmetry in the profile (Fig.~\ref{fslice}). Although the partial
double-shell morphology is not common for supernova remnants, other
remnants with similar structure have been observed \citep[see
e.g.,][]{Gaensler1998,Giacani+2009}.

We found that \GZ\ has an average surface brightness of $\Sigma_{\rm
1\, GHz} \simeq 1.4$~\Wmsr{-21}.  This SNR is therefore near the peak
of the observed distribution of $\Sigma_{\rm 1 \, GHz}$ values in the
catalogue of \citet{Green2009}, but below his estimated completeness
limit of $\simeq 1$~\Wmsr{-20}.  We note that the catalogued number of
supernova remnants in the Galaxy \citep[$n = 274$ in the catalogue of]
[]{Green2009} is lower than the $\sim$1000 generally expected from
supernova rates \citep[e.g.,][see also discussion in Brogan et al.\
2006]{TammannLS1994}, suggesting that there likely are many more
supernova remnants to be discovered at these sensitivity levels.

A relationship has been observed between the surface-brightness and
the diameter ($D$) for SNRs.  Although this $\Sigma - D$ relation has
in the past often been used to determine SNR distances it has been
shown to be only of limited value for this purpose, \citep[see,
e.g.,][]{Green2004, Green2005, BandieraP2010}.  If we take
$\Sigma_{\rm 1 \, GHz} = 1.4$~\Wmsr{-21} for \GZ\ and compare it to
the sample of 47 SNRs of known distance shown in \citet{Green2004}, we
can conclude only that the \GZ's diameter is likely between 10 and
100~pc, and its distance therefore in the not very restrictive range
of $3 \sim 30$~kpc.  The lack of apparent X-ray emission suggests high
absorption and therefore perhaps argues against the shorter end of
this range.

\subsection{\GY: A Probable \HII\ Region}
\label{sgyconc}

Our wide-field 1.4-GHz image also shows a relatively faint extended
radio source which we have called \GY.  Unlike \GZ, it does not have a
clear shell-like morphology.  Although the current data do not
reliably determine its radio spectral index, a relatively flat
spectrum, with $\alpha$ in the range of $-0.3$ to 0 is suggested (\S\
\ref{sgyobs}).  The source is present on the Glimpse $8 \, \mu$m
infrared images, and listed as G021.4571$-$00.5894 in the GLIMPSEII
source list\footnote{Available at {\ttfamily
http://irsa.ipac.caltech.edu/data/\-SPITZER/\-GLIMPSE}}. 
We therefore tentatively identify \GY\ as an \HII\ region.

\section{Summary and Conclusions}
\begin{trivlist}

\item{1.} We present a new deep 1.4-GHz radio image of the PWN
\GX\ and its environs. We also show a deep X-ray image, made from
520~ksec of \emph{Chandra} data, of the same region.

\item{2.} Although the \GX\ PWN is clearly visible in our
1.4-GHz image, we see no sign of shell radio emission from the
supernova forward shock.  We place a $3\sigma$ upper limit of
7\Wmsr{-22} on the 1-GHz surface brightness of any such emission, in
particular also on any radio emission corresponding to the
limb-brightened, non-thermal X-ray shell component.  Although low,
these limits on the radio emission are nonetheless compatible with a
broad-band model of the X-ray emission.

\item{3.} The feature called the northern knot of \GX\ is clearly seen
at 1.4~GHz. Its spectral index was $\acl = -0.04 \pm 0.31$, similar to
that of the remainder of \GX, (i.e., $\alpha \simeq 0$), although
a steeper value as is typical of shells is not excluded by the large
uncertainties.

\item{4.} We detect a new shell SNR, \GZ.  The shell has an
angular diameter of $\sim$13\arcmin.  Only part of the shell is
visible in our higher-resolution 1.4-GHz image, but it can be seen to
have a double structure with what appears as two roughly concentric
shells.  The shell structure is clearly visible on a 327-MHz image,
and the source has a total 327-MHz flux density of $\sim$2.8~Jy. The
source's spectral index is clearly non-thermal.  No $8 \, \mu$m
infrared or X-ray emission is seen.

\item{5.} A further extended radio source, \GY, is visible to
the west of \GX, which we identify as a likely \HII\ region due to the
flat radio spectral index and its detection in the infrared.

\end{trivlist}

\section*{Acknowledgements}

M. F. Bietenholz, H. C. Matheson, S. Safi-Harb and N. Bartel
acknowledge support by the Natural Sciences and Engineering Research
Council (NSERC).  S. Safi-Harb acknowledge support by the Canada
Research Chairs program. This research made use of NASA's Astrophysics
Data System.

\bibliographystyle{mn}
\bibliography{mybib1,g21temp}
\end{document}